**Description, distribution and ecology of living *Reophax pyriformis* n. sp. (Campos Basin, South Atlantic Ocean)**

Cintia Yamashita[1], Silvia Helena de Mello e Sousa[1], Michael A. Kaminski[2], Maria Virgínia Alves Martins[3,4], Carlos Eduardo Leão Elmadjian[5], Renata Hanae Nagai[6], Naira Tiemi Yamamoto[1], Eduardo Apostolos Machado Koutsoukos[7], Rubens Cesar Lopes Figueira[1]

[1]Universidade de São Paulo, Instituto Oceanográfico, Praça do Oceanográfico, 191, Cidade Universitária. 05508-120 São Paulo, SP, Brazil. smsousa@usp.br; tiemechan@gmail.com; rfigueira@usp.br

[2]Geosciences Department, King Fahd University of Petroleum & Minerals, PO Box 701, Dhahran 31261, Saudi Arabia. kaminski@kfupm.edu.sa

[3]Universidade do Estado do Rio de Janeiro, Faculdade de Geologia, Departamento de Estratigrafia e Paleontologia. Av. São Francisco Xavier, 524, sala 2020A, Maracanã. 20550-013 Rio de Janeiro, RJ, Brazil. virginia.martins@ua.pt

[4]Universidade de Aveiro, GeoBioTec, Departamento de Geociências, Campus de Santiago, 3810-193 Aveiro, Portugal.

[5]Universidade de São Paulo, Instituto de Matemática e Estatística, Rua do Matão, 1010, Cidade Universitária. 05508-090 São Paulo, SP, Brazil. elmad@ime.usp.br

[6]Centro de Estudos do Mar, Universidade Federal do Paraná, AV. Beira Mar, s/n. 83255-000 Pontal do Paraná, Paraná, Brazil. renatanagai@gmail.com

[7]Institut für Geowissenschaften, Universität Heidelberg, Heidelberg, Im Neuenheimer Feld 234, 69120 Heidelberg, Germany. ekoutsoukos@gmail.com

*Correspondence author. cintia.yamashita@usp.br

**ABSTRACT**

The distribution of living (rose Bengal stained) *Reophax pyriformis* Yamashita, Sousa and Kaminski, n. sp., an agglutinated benthic foraminiferal species, was analyzed in the area of the Campos Basin (southeastern Brazilian continental margin). The study is based on 34 oceanographic stations (54 samples), located between 400 m and 3,000 m water depth. The distribution of living *Reophax pyriformis* n.sp. density is compared to sedimentological parameters, such as total organic carbon, total nitrogen, calcium carbonate, phytopigment, lipids biomarkers (sterols, fatty acids and *n*-alcohols), total lipids, and bacterial biomass, as well as the particulate organic matter flux to the seafloor. This species was found in the range of 1,000 - 1,200 m water depth, with an average living depth of 1.52 cm in the sediment. The distribution of this species seems to be related to lipid biomarkers (allochthone and/or continental derivative, zooplankton and/or fauna, phytoplankton or primary producers) and total organic carbon under the influence of Intermediate Western Boundary Current conditions.

## 1. Introduction

The genus *Reophax* de Montfort, 1808 includes some deep infaunal (Kaminski et al., 1988, Kaminski and Gradstein, 2005; Cetean et al., 2011) and shallow infaunal forms (Koho et al., 2007; Hess and Jorissen, 2009), which are detritivores and live freely within the surface sediment (Murray, 1991; Kaminski et al., 2011). According to Kaminski et al. (2001) the absence of type material for de Montfort's (1808) species has led to the inclusion of many uniserial agglutinated taxa into the genus *Reophax*, the type species of which is *Reophax scorpiurus.*

Kaminski (1987) developed a conceptual model about the effect of sediment disturbance on a deep-sea assemblage of agglutinated foraminifera (e.g., by benthic storms, the western boundary undercurrent, turbidite deposition, and predation). In calm environments, such as the Panama Basin and Nares Abyssal Plain (with no evidence of turbidities), the agglutinated assemblages of foraminifera are dominated by organisms with branching fragile tests, such as Astrorhizidae, and species of the genus *Adercotryma*, *Reophax* and *Nodellum*. In environments with strong bottom currents, for example on the western margin of the North America Basin under the Western Boundary Undercurrent (Kaminski, 1985) or on the Hovgård Ridge in the Fram Strait (Kaminski et al., 2015), robust epifaunal and infaunal species can be found. Species of the genus *Reophax* were observed after physical disturbance, and this could be interpreted as an indicator of an environment with disturbance or fauna in the initial stage of faunistic succession (Kaminski, 1985, 1987; Kaminski and Schröder, 1987; Koho et al., 2007; Duros et al., 2011). Some *Reophax* species such as *R. excentricus* and *R. dentaliniformis* were found to colonize sediment trays placed at an abyssal site within the span of one year (Kaminski et al., 1988), while *R. scorpiurus* and *R. dentaliniformis* may indicate the presence of substrate disturbance caused by rapid sediment deposition or by strong currents (Hess and Kuhnt, 1996).

This study focuses on the Campos Basin located on the southwestern margin of the South Atlantic Ocean. The deep-sea benthic foraminiferal assemblages (especially the agglutinated foraminifera are still poorly known in this region of the world.

In this geographical context, this work aims to report and describe a new species of *Reophax*, and to analyze the occurrence and ecology of living (stained) *Reophax pyriformis* in the Campos Basin.

## 2. Study area

The study area is located between 21°S and 23°S and 38°W and 42°W in the Campos Basin on the southeastern Brazilian continental slope and São Paulo Plateau (Figure 1; Viana et al., 1998a,b). The São Paulo Plateau is an extension of the continental slope that progresses structurally up to the Vitória-Trindade Chain, and is located between 2,900 m and 3,200–3,400 m deep (Zembruski, 1979; Almeida and Kowsmann, 2016).

Three currents are present in the Campos Basin: the Brazil Current (BC), the Intermediate Western Boundary Current (IWBC), and the Deep Western Boundary Current (DWBC) (Silveira et al., 2017). The BC is observed from the surface down to intermediate waters and flows south-southwestward, reaching speeds up to 0.80 $m.s^{-1}$ and transporting the Tropical Water (TW) and the South Atlantic Central Water (SACW).

The IWBC occurs from 500 m to 1,200 m water depth and carries the Antarctic Intermediate Water (AAIW) with a contribution of the Upper Circumpolar Water (UCW) to north-northeastward (Figure 2). This current has a swift core of velocities centered at 800 m deep that exceeds 0.30 $m.s^{-1}$ (Böebel et al., 1999; Silveira et al., 2004). Below 2,000 m depth, the DWBC flows south-southwestward and transports the North Atlantic Deep Water (NADW) with a contribution of the lower UCW. The presence of the São Paulo Plateau shifts the DWBC core offshore, which reduces the interaction of the western boundary current system (BC and IWBC) with the DWBC (Sousa et al., 2006). The Brazil Current System is characterized by meanders and eddies, including those formed by the BC and IWBC (Mascarenhas et al., 1971; Signorini, 1978; Campos et al., 2000; Palóczy et al., 2014).

The Campos Basin is considered a meso-oligotrophic system (Burone et al., 2011; Suzuki et al., 2015). The area is characterized by the occurrence of temporary coastal upwelling, shelf break upwelling, coastal fronts, and meanders and eddies associated with the instability of the BC (Castro and Miranda, 1998; Palóczy et al., 2014).

In the southeast Brazilian coast, the chlorophyll-*a* concentrations vary from 0.2 to 3.0 $mg.m^{-3}$ in the areas influenced by the hydrodynamic features like eddies of BC and shelf edge and coastal upwelling of the SAWC and by the relatively limited input of rivers (Marone et al., 2010).

In the study area, the sedimentary cover of the upper slope was determined to consist of fine sand or sandy mud. The middle slope (550–1,200 m) is characterized by iron-rich laminated indurated fine sands and deep-water coral mounds (Caddah et al., 1998; Viana et al., 1998b). Below 1,200 m water depth, a thin (<10 cm thick) Holocene calcareous ooze (a mixed coccolithophore/nannoplankton-foraminiferal ooze) is observed overlying the iron-rich crust (Viana et al., 1998b).

## 3. Materials and methods

Sediment samples collected at 34 oceanographic stations, arranged along five transects (approximately between the 400 m and 3,000 m isobaths) in the Campos Basin (Brazil) were considered in this study (Figure 1, Additional Data). The samples were collected during the austral summer of 2009, using the research vessels R/V Gyre and Miss Emma McCall.

Sediment samples were collected using a TDI-Brooks box corer (50 x 50 x 50 cm). At each station, replicate cores (10 cm diameter x 20 cm height) were taken from the box core collected for each parameter (geochemical, bacterial biomass, and living foraminifera). The upper 0–2 cm of the cores were sliced to be analyzed for geochemical parameters (calcium carbonate, total organic carbon, total nitrogen, and total lipid contents), bacterial biomass, and living foraminifera. At stations I06 and I08 (pseudoreplicates), the sediment core was sliced into 1 cm thick slices in a 0–10 cm interval (see details in Ribeiro-Ferreira et al., 2017a).

### 3.1 Benthic foraminifera

For foraminiferal analyses, an aliquot of 50 $cm^3$ of sediment was treated and stained with rose Bengal (1 g of rose Bengal in 1,000 ml of alcohol) to differentiate between living and dead foraminifera (Walton, 1952). In order to avoid the fragmentation of foraminifera, the wet samples (aliquot of 50 $cm^3$) were gently washed through 125 µm and 63 µm sieves. In this work, only the absolute abundance (n.º ind./50 $cm^3$ in the sediment fraction >63 µm) of living specimens of *R. pyriformis* n.sp. is reported.

In order to describe the vertical distribution of individual taxa, we used the average living depth (ALD) after Jorissen et al. (1995). The ALD is calculated with the following formula:

$$ALD_x = \sum_{i=1,x}^{x} \frac{n_i D_i}{N} \qquad (1)$$

where, $x$= lower boundary of the deepest sample;

$n_i$ = number of individuals in interval i;

$D_i$ = midpoint of sample interval i;

$N$ = total number of individuals for all levels.

For the station I06 and I08, $ALD_{10}$ was calculated for *R. pyriformis* n.sp., on the basis of the numbers of stained individuals found in successive sediment slices.

Selected specimens were imaged using a ZEISS Camera AxioCam ICc3 Rev.3 and a digital Scanning Electron Microscope (SEM) QUANTA FEG 650. These specimens were taken from stations I09 and H08 (Appendix A; Figure 3). The elemental composition of the tests of three specimens of *R. pyriformis* n.sp. was analyzed with Energy-dispersive X-ray spectroscopy (EDX), using an SEM. In addition, the elemental composition of some points/small areas in the tests of the selected specimens were analyzed. The surface of the foraminiferal tests was coated with platinum and was not polished.

A three-dimensional model of the species was created using the data provided by tomography to see the internal arrangement of the chambers, which consisted of 571 grayscale slices with a resolution of 456 x 456 pixels of one specimen of *R. pyriformis* n. sp. Each one of these slices was subjected to a preprocessing step, in which a median filter with a 5 x 5 kernel was applied in order to reduce noise, but also to preserve foreground edges. An entropy-based thresholding algorithm (Yen et al., 1995) was used so that a binary criterion could be established to distinguish what was part of the species and what was not. Following this step, a 3D surface was created using the 3D Viewer plugin from ImageJ software through a procedure of connecting foreground edge-located pixels by the shortest possible segment between neighboring slices. Since the resulting 3D model still presented many artifacts and sharp edges that could be regarded as defects, a manual refinement was applied subsequently to the mesh using the Blender computer graphics toolset. This software also allowed the improvement of the model quality by smoothing the edges created between slices and thickening the interior walls of the sample. An example is shown in Figure 4.

**3.2 Bacterial data, Sedimentological data and Vertical flux estimation of particulate organic matter**

The bacterial data, calcium carbonate content ($CaCO_3$), total organic carbon (TOC), lipid biomarkers, the phytopigment (chlorophyll *a* + pheophytin *a*), and particulate organic flux to the seafloor were already analyzed by Yamashita et al. (2018a).

Synthetically, the nucleic acids (DNA and RNA) were stained with 2.5 μM of the Syto13 fluorochrome (Molecular Probes, ref S-7575) before determining the total bacterial abundance by a CyAn ADP (DakoCytomation) cytometer. The abundance data obtained in this work was dimensioned as the number of cells or bacteria per wet sediment mass (standardized per gram). Total prokaryotic cell abundance was based on stained cells and fluorescent sphere. The bacterial biomass was calculated by a conversion factor of $20.10^{-15}$ g per cell (Lee and Fuhrman, 1987).

Calcium carbonate content ($CaCO_3$) was determined by the difference in weight of the sediment prior to and after acidification of each sample with 1.0N HCl (Mahiques et al. 2004).

Total organic carbon (TOC) was determined by a CHNS/O Perking Elmer analyzer (2400 series II) (see details in Ribeiro-Ferreira et al., 2017b).

Lipid biomarkers were determined following a published method (Oliveira et al., 2012). Sterols and *n*-alcohols in the neutral fraction were identified and quantified by gas chromatography/mass spectrometry as TSM-derivatives. Fatty acids in the acidic fraction of the bulk extracted were methylated ($BF_3$/MeOH at 85ºC for 2h) and determined as fatty acid methyl esters (FAMEs) by gas chromatography with flame ionization detection (see details in Ribeiro-Ferreira et al., 2017b).

The phytopigment (chlorophyll *a* + pheophytin *a*) was determined by a UV-Vis Perkin-Elmer® Lambda 20 spectrophotometer and Turner Designs® TD-700 Fluorimeter. The calibration was performed with pure chlorophyll *a* (Sigma® C-6144) (see details in Ribeiro-Ferreira et al., 2017b).

The particulate organic flux to the seafloor was determined based on the Dunne et al. (2005) models. The models were implemented with satellite data (MODIS and SeaWiFS). To determine the fraction of nanoplankton and picoplankton, the model proposed by Ciotti et al. (2002) was applied.

### 3.3 Data analysis

Spearman correlation analyses (non-parametric data) were performed considering $p<0.05$ as the significant level. The density of *R. pyriformis* n.sp. (n.º ind./50 $cm^3$) was correlated with organic carbon, total nitrogen, total organic carbon, calcium carbonate content, bacterial biomass, phytopigment, lipid biomarkers (total concentration and grouped as distinct sources of OM in terrigenous, zooplankton/fauna, primary producers, and bacteria), total lipids, and particulate organic matter flux to the seafloor using STATISTICA version 10.

## 4. Results

### 4.1 Systematic taxonomy

Phylum FORAMINIFERA d'Orbigny, 1826

Class GLOBOTHALAMEA Pawlowski, Holzmann & Tyszka, 2013

Subclass TEXTULARIIA Mikhalevich, 1980

Order LITUOLIDA Lankester, 1885

Suborder HORMOSININA Haeckel, 1894

Superfamily HORMOSINOIDEA Haeckel, 1894

Family REOPHACIDAE Cushman, 1927

Genus Reophax Montfort, 1808

*Reophax pyriformis* Yamashita, Sousa and Kaminski n. sp.

Plate 1, Figs. A-D

*Description:* Test free, uniserial, small, arcuate, comprised of four to seven pyriform chambers, tapering towards the aperture. Chambers increase in size slowly. The wall is coarsely agglutinated, with mineral grains (mostly quartz) and biogenic fragments such as sponge spicules and fragments of carbonate shell, mostly protruding from the test wall, obscuring the chamber shape and the sutures. Aperture terminal, a small round opening on a produced, finely-agglutinated neck.

*Remarks:* Differs from *Reophax caribensis* Seiglie & Bermúdez, 1969 in possessing pyriform chambers. Differs from *Reophax pyrifera* Rhumbler, 1905, in its smaller dimensions and its coarsely agglutinated wall.

*Type Locality*: Southwestern Atlantic, offshore Brazil. The holotype is from sample I09 (21,1858°S, 40,0523°W) collected at 1,300 m water depth.

*Type Level:* Recent.

*Dimensions:* Length of holotype: 551 µm; Length of final chamber: 147 µm.

Type specimens: Holotype (Plate 1A) is deposited in the micropaleontological collections of "Prof. Edmundo F. Nonato Biological Collection" at the "Instituto Oceanográfico" of "Universidade de São Paulo", catalog numbers For-00002. Additional unfigured specimens are housed in the European Micropalaeontological Reference Center, Micropress Europe, Kraków, Poland. Other paratypes are deposited in the Laboratory of Environmental Bioindicators at the University of São Paulo, Brazil (Plate 1B, 1C and 1D).

*Material:* 89 specimens from 11 samples.

*Derivation of name:* From the tapering, pyriform shape of the chambers.

*Occurrence*: The sum of the number of individuals of *R. pyriformis* was <117 ind. in all stations (Appendix A). This species was recorded in 11 of 54 samples in the Campos Basin (Figure 1). It was observed in the middle continental slope (1,000 to 1,200 m water depth). It was also observed in the analyzed depth range of the bottom sedimentary column between 0 and 10 cm (Appendix A).

**4.2 Average Living Depth**

No specimens of *R. pyriformis* were observed at station I06 (400 m water depth) in the analyzed depth range of the sedimentary column. At station I08 (993 m water depth) the $ALD_{10}$ was 1.52 cm for this species (Appendix A).

**4.3 3D-model**

The tomographic images are presented in Figure 4 showing external (Figure 4A) and internal (Figures 4 B, C and D) features of the test. It is possible to see four tight pyriform chambers.

**4.4 EDX Analysis on the SEM**

The results of the elemental composition of three different specimens of *R. pyriformis* are presented in Figure 3. The results of the composition of the test wall of this species mainly indicate the presence of silicon and oxygen in all analyzed specimens. Minor amounts of Fe, Ca, and Al were detected in some specimens.

### 4.5 Environmental factors

The results of biotic and abiotic variables of all the analyzed stations in this work are reported in Yamashita et al. (2018b) (Appendix A). The stations with *R. pyriformis* presented the following range of values: 29.13–37.7% for $CaCO_3$; 10.43 mg.g$^{-1}$ to 16.30 mg.g$^{-1}$ for TOC; 5.90–19.63 μg-g$^{-1}$ for total lipids (i.e., sum of sterols, n-alcohols and identified fatty acids); 20.27–63.60 mg C m$^{-2}$ day$^{-1}$ for the particulate OM flux and 0.44-1.10 mg C.g$^{-1}$ for the bacterial biomass. The lipid biomarkers (sterols, n-alcohols and FAMEs) were grouped according to potential sources of OM (Oliveira et al., 2012).

The biomarker concentrations varied: allochthone and/or continental between 0.42 μg.g$^{-1}$ and 1.81 μg.g$^{-1}$; zooplankton and/or fauna between 0.94 μg.g$^{-1}$ and 3.31 μg.g$^{-1}$; phytoplankton and/or primary producers between 1.63 10$^{-1}$ μg.g$^{-1}$ and 7.53 μg.g$^{-1}$; bacteria between 0.16 μg.g$^{-1}$ and 0.47 μg.g$^{-1}$.

Correlation of *R. pyriformis* n.sp. density (ind./50cm$^3$), abiotic and biotic variables is presented in Table 1. *Reophax pyriformis* n.sp. density exhibited a moderate and positive significant correlation with total organic carbon (0.40), total lipids (0.40) and weak and positive correlations with allochthone and/or continental derivative (0.35), zooplankton and/or fauna (0.37), and phytoplankton or primary producers (0.36).

## 5. Discussion

### 5.1 *Reophax pyriformis* n.sp.

*Reophax pyriformis* n.sp. is relatively selective in the choice of detrital materials for the construction of its test. The EDX results, revealing the presence of high Si and O contents, suggest that this species constructs its test mainly with quartz grains (Heron-Allen and Earland, 1909; Martins et al., 2016; Yamashita et al., 2018a,b).

It should be noted that there is variability in the number of chambers in the analyzed specimens.

**5.2 Distribution of *Reophax pyriformis* n.sp. on the Campos Basin continental slope**

In the Campos Basin, *R. pyriformis* n.sp. was found at stations of the continental slope, mostly between 1,000–1,200 m of water depth. According to Kaminski et al. (1988), Murray (1991), and Hess and Kuhnt (1996) the genus *Reophax* is infaunal, free living, and a detritivore. The $ADL_{10}$ value of 1.52 cm estimated for *R. pyriformis* n.sp. (observed living between 0 and 10 cm in the sedimentary column) in the Campos Basin, allows to consider it as a free living and shallow infaunal species according to the definition of Rathburn and Corliss (1994) for infaunal microhabitats.

In general, oxygen concentration and food availability are considered to be limiting factors for the establishment of living benthic foraminifera communities in deep sea environments (Kaiho 1991, 1994; van der Zwaan, 1999; Alve and Bernhard, 2003; Gooday, 2003). However, on the Campos Basin slope, oxygen concentration does not seem to be a limiting factor for the establishment of benthic foraminifera, as the availability of food seems to be the most important factor governing the distribution of foraminiferal assemblages along the Campos Basin oligotrophic continental slope (Sousa et al., 2017; Yamashita et al., 2018a,b).

According to Yamashita et al. (2018a,b), in the Campos Basin, the Brazil Current System controls the sediment phytopigment concentrations, which plays an important role in the distribution of foraminifera. However, no significant positive correlations were observed between *R. pyriformis* and phytopigment concentrations, neither with particulate organic matter vertical flux, nor bacterial biomass (Table 1). Significant correlations were found between this species and total organic carbon, lipid biomarkers (allochthone and/or continental derivative, zooplankton and/or fauna, phytoplankton or primary producers), and total lipids in the sediments. This can be an indication that *R. pyriformis* n.sp. has an opportunistic and detritivore behavior.

Additionally, along the continental slope of the Campos Basin, the bottom morphology may induce changes in velocities of currents (Viana, 2002). These velocities also change as a function of depth and decrease as the distance from the core region of the currents increases (Silveira et al., 2017). In the Campos Basin, *Reophax pyriformis* n.sp. was observed (Figure 1) mainly under currents with velocities of about 0.3 $m.s^{-1}$, and in the depth range of the IWBC. This type of distribution supports the idea that this species can tolerate the disturbance caused by sediment remobilization and transport. The increase in this species density (Appendix A) may indicate a region with physical disturbances (for example by contour currents) (Kaminski, 1985, 1987; Yamashita et al., 2018a,b), as observed in the São Paulo Bight, South Atlantic (Mahiques et al., 2017), and on the continental rise off Nova Scotia (North Atlantic) (Kaminski, 1985).

## 6. Conclusions

*Reophax pyriformis* n.sp. is described for the first time in this study along with some characteristics of its ecology. In the Campos Basin, *R. pyriformis* occurs at 1,000–1,200 m of water depth as a free living shallow infaunal species. Its distribution is related to relatively high contents of total organic carbon and the following items indicated by lipid biomarkers in the sediments: allochthone and/or continental derivative, zooplankton and/or fauna, and phytoplankton or primary producers. This supports the assumption that this species has an opportunistic and detritivore behavior. *Reophax pyriformis* seems to feed on all kind of sedimentary organic matter present on the continental slope of the Campos Basin under the influence of the IWBC.

**Acknowledgments**

The parameters for the organic carbon and total nitrogen were kindly provided by Dr. Carlos E., Rezende (*Universidade Estadual Norte Fluminense*), the particulate organic matter vertical flux by Thaisa Vicente (*Fundação para Desenvolvimento Tecnológico da Engenharia*), the lipids biomarkers by Dr. Renato Carreira (*Pontíficia Universidade Católica do Rio de Janeiro*), the bacterial biomass by Dr. Rodolfo Paranhos (*Universidade Federal do Rio de Janeiro*). The authors are indebted to MSc. Nancy Taniguchi, Dr. Mônica Petti and Dr. Sueli Godoi (*Universidade de São Paulo*) who are gratefully acknowledged for technical support. We also thank the anonymous reviewers and Dr. Sibelle Trevisan Disaró for their valuable comments and criticism.

**Funding sources**

We are grateful to PETROBRAS for providing sediment samples and financial support under the "Heterogeneidade Ambiental da Bacia de Campos" project, especially to Ana Falcão and Helena Lavrado. This work was also supported by *Fundação de Amparo à Pesquisa do Estado de São Paulo* (FAPESP) under "*Avaliação da produtividade oceânica no talude continental da Bacia de Campos, margem sudeste Brasileira: uma visão a partir de foraminíferos bentônicos e isótopos estáveis*" project (Proc. n° 2013/12510-3), *"Distribuição vertical dos foraminíferos bentônicos vivos dos mud-belts na margem continental sul brasileira: resposta à variação da condição redox e ao aporte de matéria orgânica"* project (Proc. n° 2017/00427-5), and *Conselho Nacional de Desenvolvimento Científico e Tecnológico* (n° 164640/2013-4). MAK thanks the Deanship of Scientific Research, King Fahd University of Petroleum & Minerals, for support under Project IN161002.

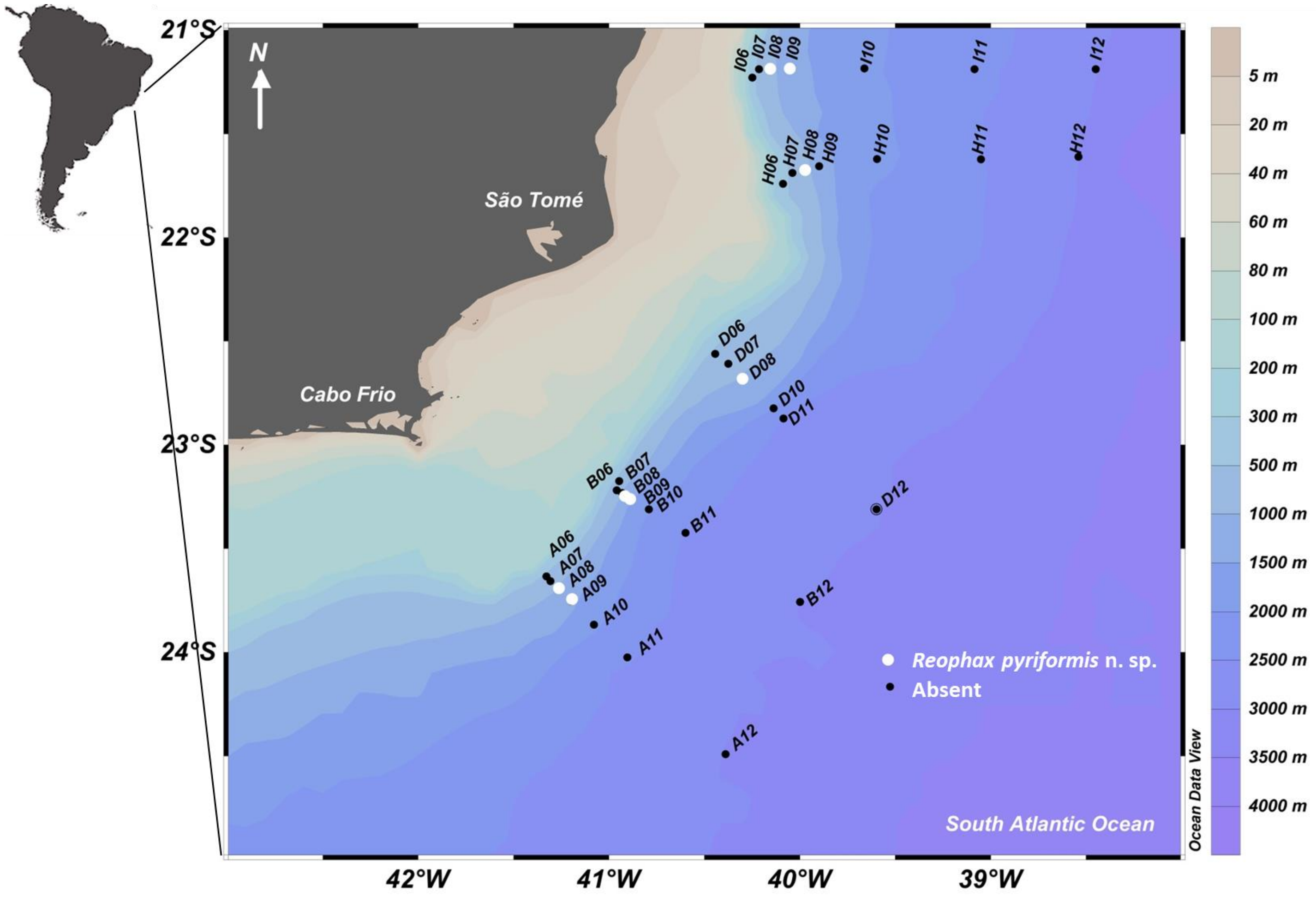

Figure 1. Location of Campos Basin-sampling sites and the occurrence of *Reophax pyriformis* n. sp. (Color online only).

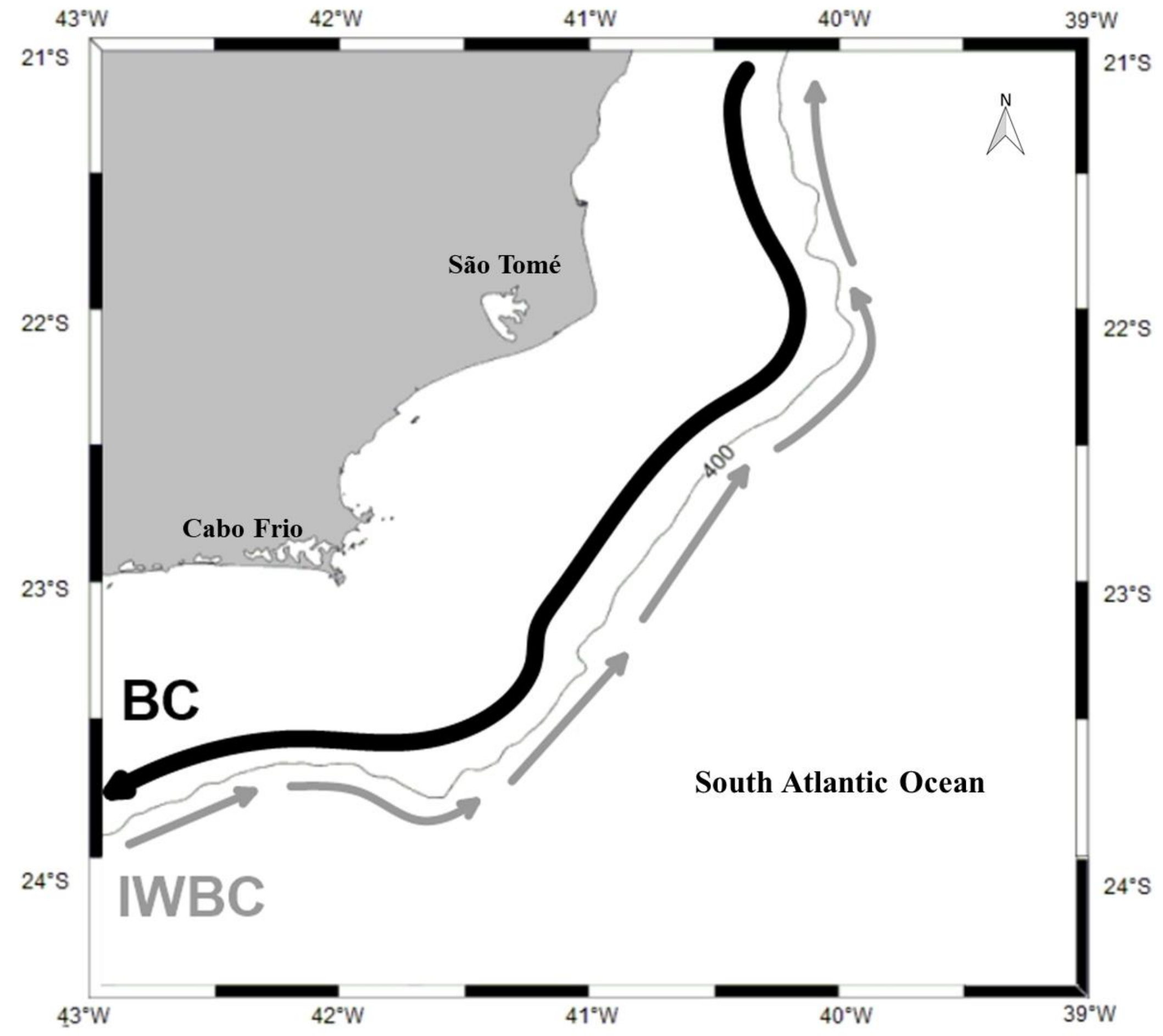

Figure 2. Schematic representation of Brazil Current (BC) and Intermediate Western Boundary Current (IWBC) in Campos Basin. Modified from Yamashita et al. (2018b)

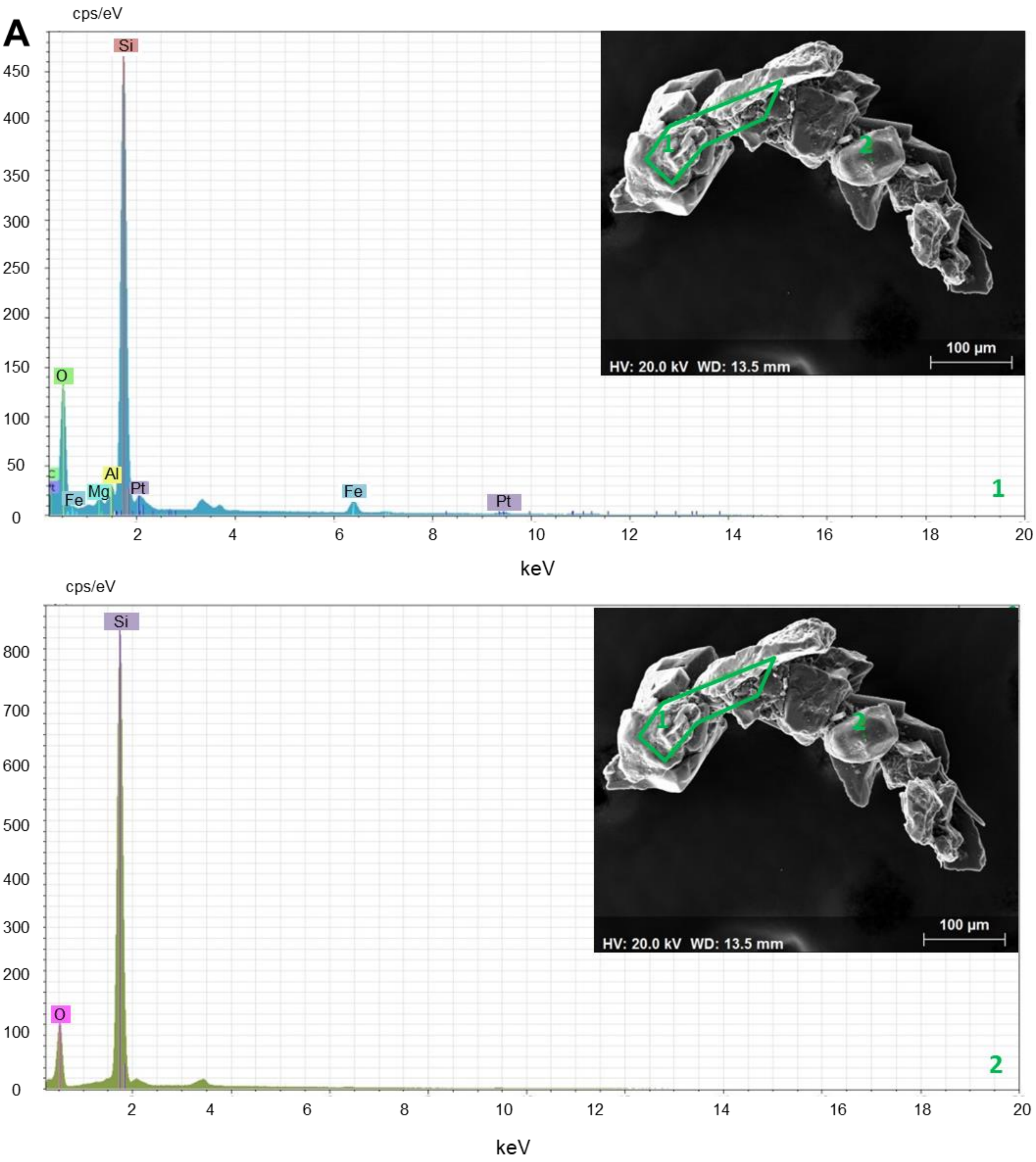

B

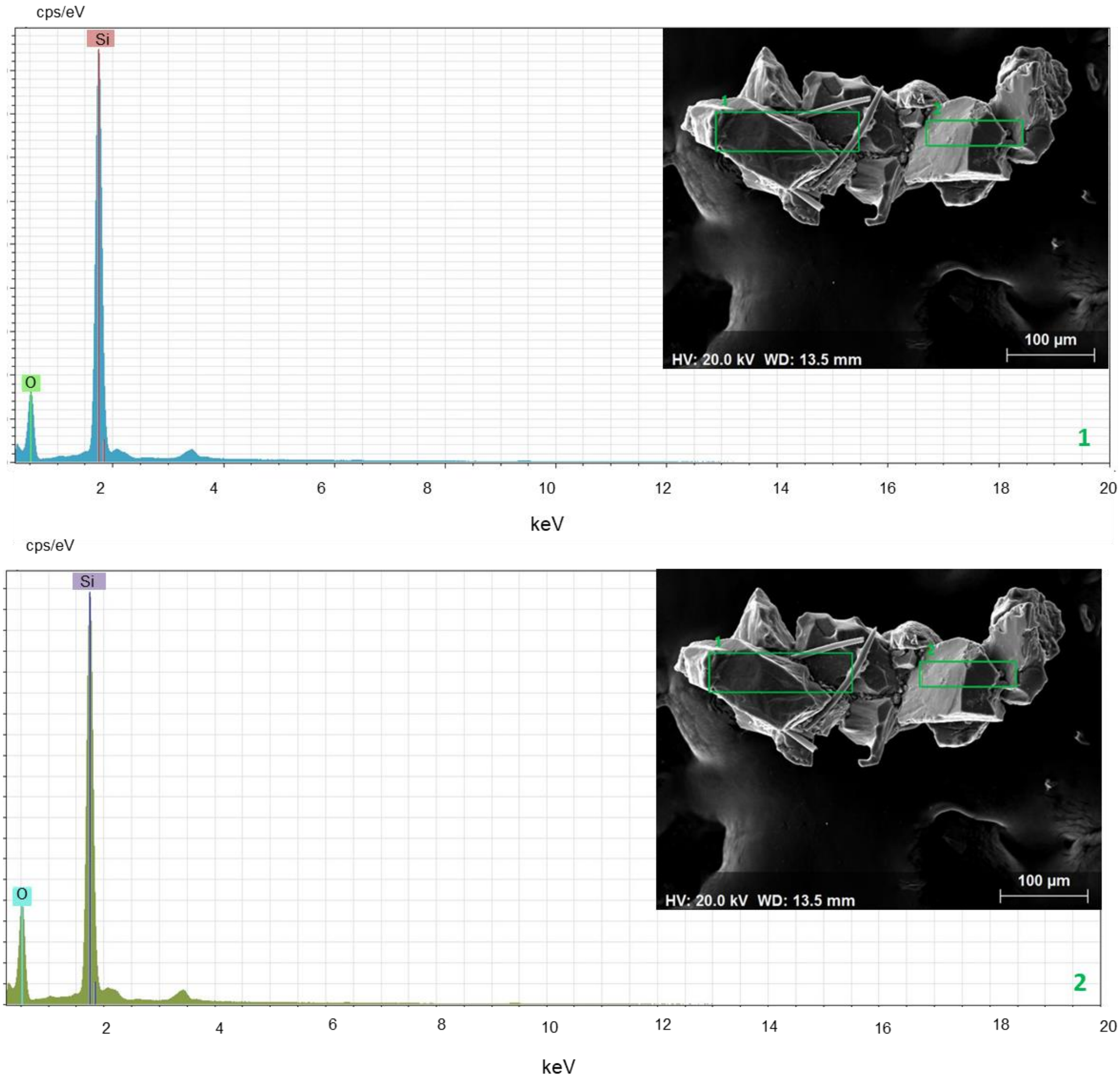

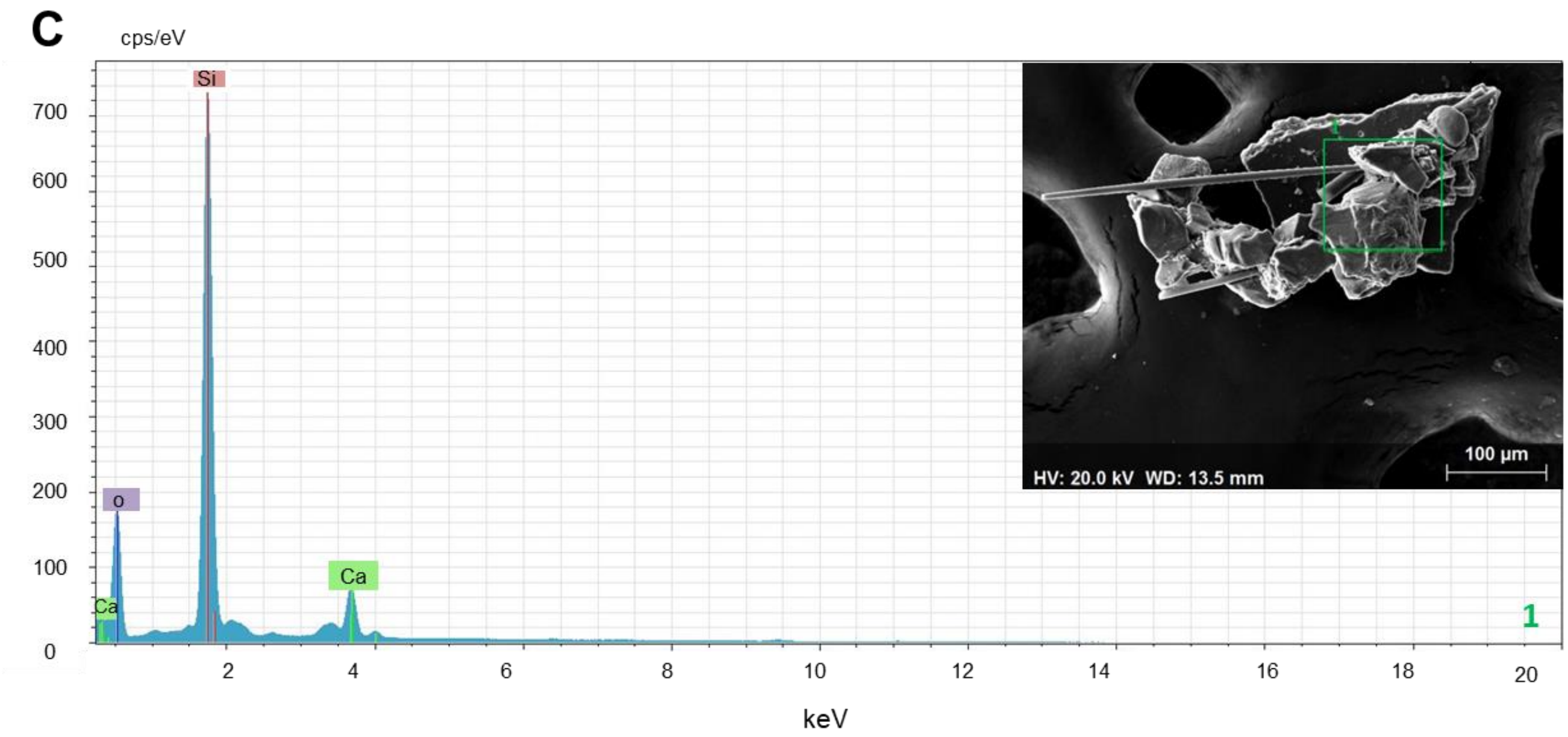

Figure 3. Results of the three EDX analyses on SEM with general elemental composition of the test of *Reophax pyriformis* n. sp. (A and C from I09 0-2cm; and B from H08 0-2cm; see Latitude and longitude in the Appendix A). The elements Silicon (A, B, C); Oxygen (A, B, C); Calcium (A, C), iron (A), aluminum (A) and Platinum (A) were identified in the test composition of this species (Color online only).

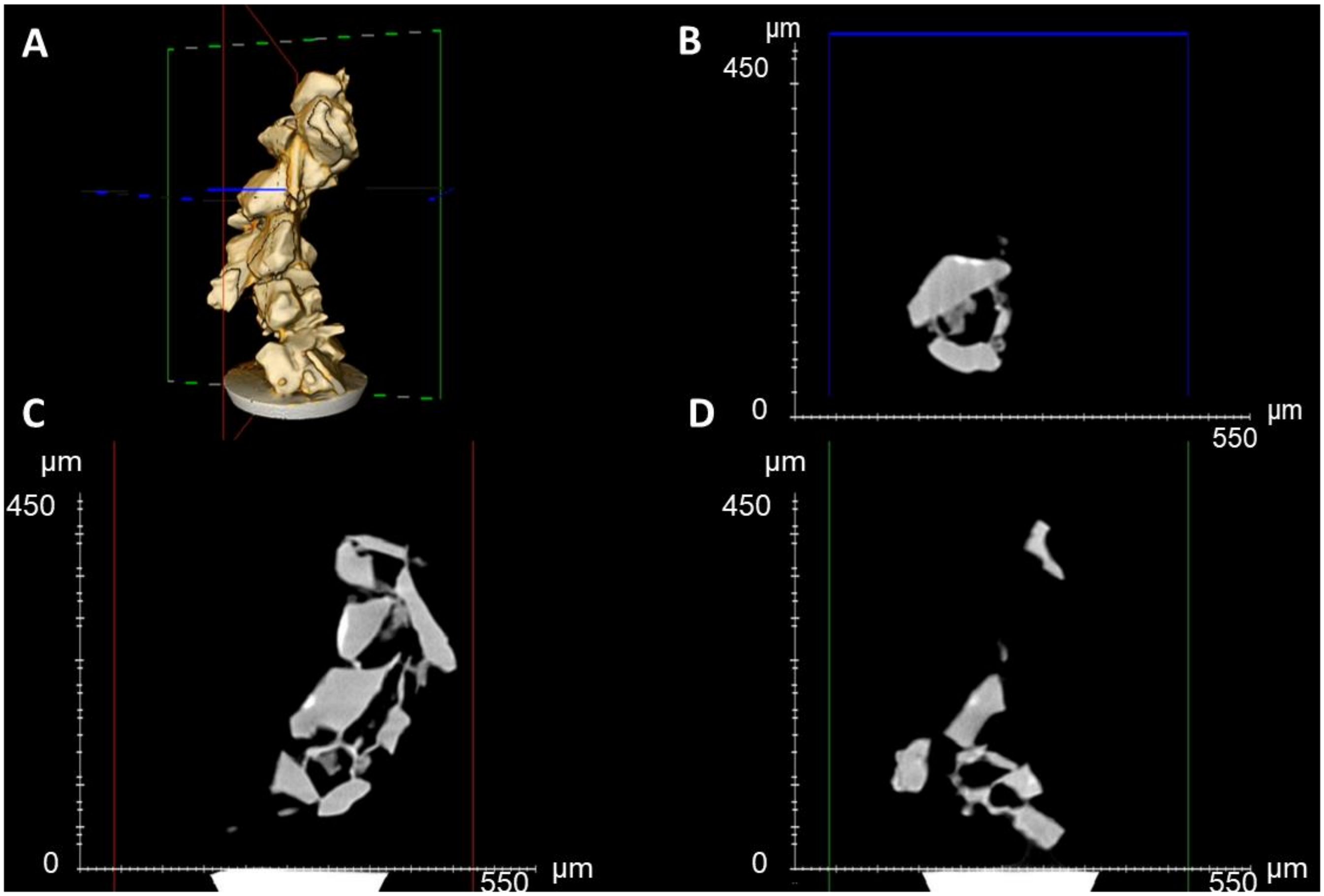

Figure 4. 3D model of the *Reophax pyriformis* n. sp. (station I08). Colored rectangles indicate the cutting plane (red, blue, and green).A- show its external features, while planes B, C and D show internal features. B - shows the transversal section; C – shows the arrangement of the chambers; C and D – together these images show longitudinal sections evidencing the pyriform chambers. (Color online only).

Plate 1. Images of living (rose Bengal stained) specimens of *Reophax pyriformis* n. sp. taken in SteREO Discovery.V20 (A) and Scanning Electron Microscope (B, C and D): A-station I09 (curatorial museum numbers For-00002); B- station H08; C- station I09 (curatorial museum numbers For-00003); D- station I08 (curatorial museum numbers For-00004) (Color online only).

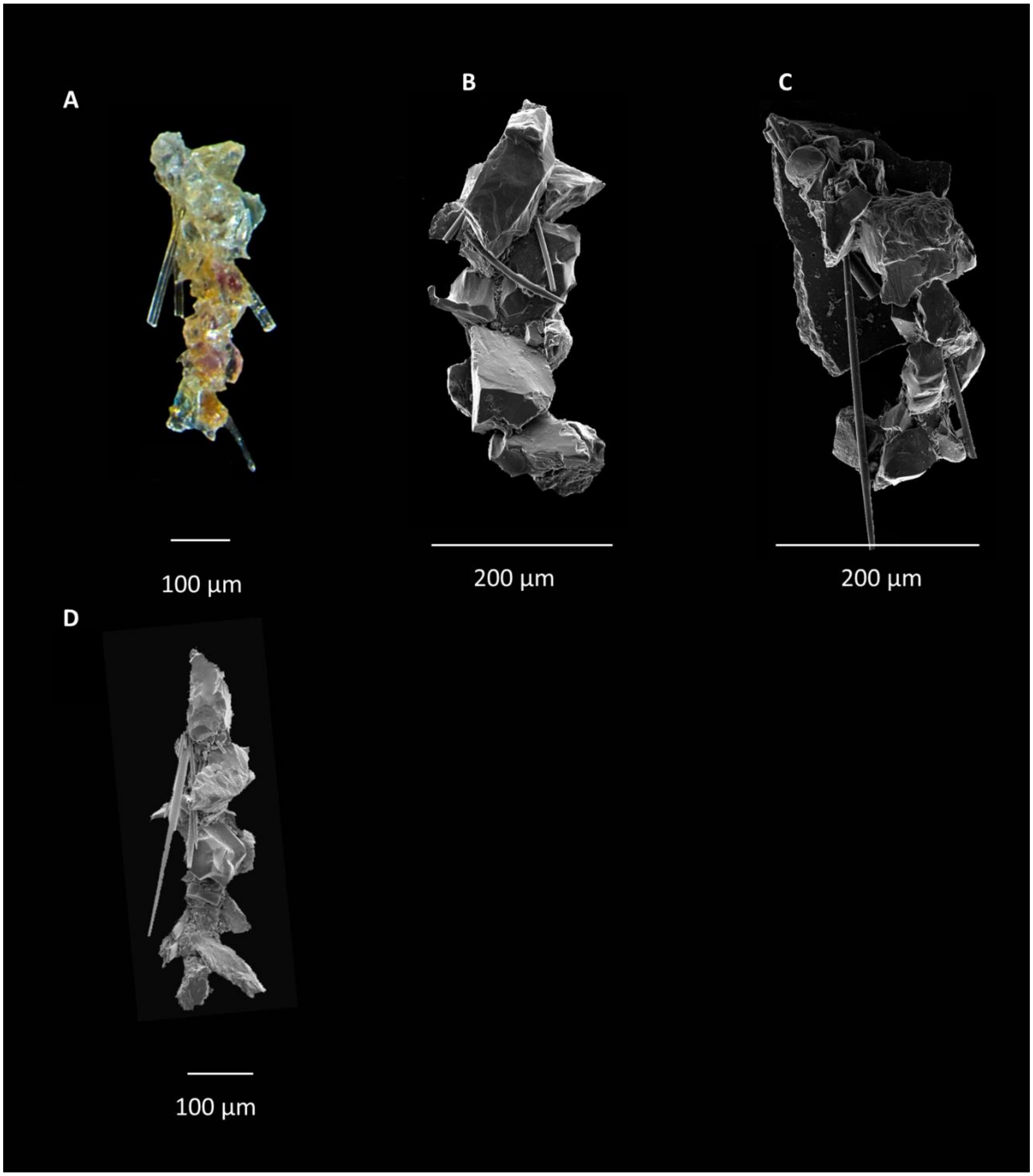

**Table 1**: Correlation of *R. pyriformis* n.sp. density (ind./50cm$^3$), abiotic and biotic variables. Bold values indicate significant correlations at $p<0.05$.

| | ***R. pyriformis* n.sp. (ind./50cm$^3$)** |
|---|---|
| **Bacterial biomass (mgC.g$^{-1}$)** | 0.27 |
| **Total organic carbon (mg.g$^{-1}$)** | **0.40** |
| **Calcium carbonate content (%)** | **-0.50** |
| **Phytopigment (µg.g$^{-1}$)** | 0.28 |
| **Vertical particulate organic flux (mgC.m$^{-2}$.day$^{-1}$)** | 0.18 |
| **Allochthone and/or continental derivative (µg.g$^{-1}$)** | **0.35** |
| **Zooplankton and/or fauna (µg.g$^{-1}$)** | **0.37** |
| **Phytoplankton or primary producers (µg.g$^{-1}$)** | **0.36** |
| **Bacteria (µg.g$^{-1}$)** | 0.34 |
| **Total lipids (µg.g$^{-1}$)** | **0.40** |

## Appendix A. Supplementary data

Geographical coordinates of the analyzed stations (WGS84), water depth, and density of living specimens of *R. pyriformis* n. sp. (number of individuals/50cm$^3$) and biotic and abiotic variables. Where: Biomass= Bacterial biomass (mg C.g$^{-1}$); TOC= Total organic carbon (mg.g$^{-1}$); $CaCO_3$=Calcium carbonate content (%); Phytopigment (µg.g$^{-1}$), Flux=Vertical particulate organic flux (mgC.m$^{-2}$.day$^{-1}$); Terr= Alloctone and/or continental derivative biomarker (µg.g$^{-1}$); Zoo= Zooplankton and/or fauna biomarker (µg.g$^{-1}$); PP= Phytoplankton and/or primary producers biomarker (µg.g$^{-1}$); Bacteria biomarker (µg.g$^{-1}$); Lipids (total lipids; µg.g$^{-1}$)*= no data.

| **Station** | **Depth (m)** | **Lat. (WGS84)** | **Long. (WGS84)** | **N (ind./50cm$^3$)** | **Biomass (mg C.g$^{-1}$)** | **TOC (mg.g$^{-1}$)** | **$CaCO_3$ (%)** | **Phytopigment (µg.g$^{-1}$)** | **Flux (mgC.m$^{-2}$.day$^{-1}$)** | **Terr (µg.g$^{-1}$)** | **Zoo (µg.g$^{-1}$)** | **PP (µg.g$^{-1}$)** | **Bacteria (µg.g$^{-1}$)** | **Lipids (µg.g$^{-1}$)** |
|---|---|---|---|---|---|---|---|---|---|---|---|---|---|---|
| **A06 (0-2 cm)** | 391 | -23,6331 | -41,3288 | 0 | 1.70 | 7.37 | 35.30 | 4.03 | 38.64 | 0.66 | 1.56 | 2.54 | 0.29 | 8.11 |
| **A07 (0-2 cm)** | 699 | -23,6560 | -41,3083 | 0 | 1.51 | 15.63 | 31.00 | 3.83 | 38.47 | 1.41 | 3.08 | 4.01 | 0.42 | 14.71 |
| **A08 (0-2 cm)** | 1018 | -23,6860 | -41,2684 | 5 | 0.74 | 14.33 | 30.59 | 3.65 | 27.20 | 1.21 | 2.70 | 5.16 | 0.45 | 19.24 |
| **A09 (0-2 cm)** | 1319 | -23,7527 | -41,1981 | 2 | 1.06 | 10.43 | 33.44 | 2.69 | 20.27 | 0.42 | 0.94 | 1.63 | 0.26 | 10.40 |
| **A10 (0-2 cm)** | 1935 | -23,8664 | -41,0792 | 0 | 0.47 | 9.80 | 42.18 | 0.57 | 10.46 | 0.34 | 0.64 | 0.60 | 0.15 | 6.44 |
| **A11 (0-2 cm)** | 2493 | -24,0239 | -40,9043 | 0 | 0.23 | 6.67 | 56.81 | 0.17 | 9.33 | 0.44 | 0.49 | 0.55 | 0.20 | 3.65 |
| **A12 (0-2 cm)** | 3035 | -24,4900 | -40,3902 | 0 | 0.19 | 5.20 | 71.03 | 0.15 | 9.27 | 0.17 | 0.20 | 0.16 | 0.07 | 1.32 |
| **B06 (0-2 cm)** | 412 | -23,1734 | -40,9471 | 0 | 0.73 | 20.50 | 31.16 | 5.92 | 90.54 | 2.11 | 2.60 | 5.47 | 0.60 | 18.11 |
| **B07 (0-2 cm)** | 738 | -23,2176 | -40,9609 | 0 | 1.60 | 20.73 | 31.15 | 10.19 | 43.79 | 1.98 | 6.54 | 9.33 | 0.63 | 27.17 |
| **B08 (0-2 cm)** | 1001 | -23,2307 | -40,9320 | 11 | 0.94 | 16.30 | 31.29 | 4.78 | 26.71 | 1.81 | 3.31 | 7.53 | 0.43 | 19.63 |
| **B09 (0-2 cm)** | 1228 | -23,2537 | -40,8986 | 4 | 1.03 | 15.80 | 29.13 | 3.16 | 22.05 | 1.74 | 1.72 | 3.29 | 0.47 | 14.01 |

# APPENDIX A

Continued

| Station | Depth (m) | Lat. (WGS84) | Long. (WGS84) | N (ind./50cm$^3$) | Biomass (mg C.g$^{-1}$) | TOC (mg.g$^{-1}$) | $CaCO_3$ (%) | Phytopigment (µg.g$^{-1}$) | Flux (mgC.m$^{-2}$.day$^{-1}$) | Terr (µg.g$^{-1}$) | Zoo (µg.g$^{-1}$) | PP (µg.g$^{-1}$) | Bacteria (µg.g$^{-1}$) | Lipids (µg.g$^{-1}$) |
|---|---|---|---|---|---|---|---|---|---|---|---|---|---|---|
| **B10 (0-2 cm)** | 1900 | -23,3104 | -40,7914 | 0 | 0.62 | 11.80 | 39.07 | 1.02 | 15.33 | 0.81 | 0.92 | 1.41 | 0.11 | 4.91 |
| **B11 (0-2 cm)** | 2492 | -23,4226 | -40,5998 | 0 | 0.26 | 7.53 | 58.93 | 0.26 | 11.26 | 0.49 | 0.54 | 0.69 | 0.20 | 3.96 |
| **B12 (0-2 cm)** | 2424 | -23,7556 | -39,9999 | 0 | 0.28 | 4.40 | 58.48 | 0.10 | 9.39 | 0.14 | 0.09 | 0.12 | 0.01 | 0.70 |
| **D06 (0-2 cm)** | 401 | -22,5603 | -40,4442 | 0 | 0.76 | 10.63 | 36.77 | 2.94 | 44.90 | 1.17 | 1.61 | 2.47 | 0.29 | 9.14 |
| **D07 (0-2 cm)** | 696 | -22,6075 | -40,3756 | 0 | 0.66 | 11.70 | 34.38 | 2.34 | 32.10 | 0.84 | 1.08 | 1.76 | 0.21 | 6.22 |
| **D08 (0-2 cm)** | 1010 | -22,6829 | -40,2942 | 7 | 0.44 | 13.20 | 36.03 | 1.53 | 28.80 | 1.43 | 2.50 | 3.99 | 0.32 | 12.68 |
| **D10 (0-2 cm)** | 1921 | -22,8231 | -40,1386 | 0 | 0.20 | 7.23 | 48.41 | 0.72 | 13.00 | 0.23 | 0.52 | 0.68 | 0.10 | 2.43 |
| **D11 (0-2 cm)** | 2492 | -22,8713 | -40,0865 | 0 | 0.22 | 3.25 | 65.48 | 0.64 | 11.90 | 0.31 | 0.26 | 0.41 | 0.04 | 1.47 |
| **D12 (0-2 cm)** | 3016 | -23,3101 | -39,5992 | 0 | 0.19 | 4.17 | 73.86 | 0.14 | 9.10 | 0.28 | 0.20 | 0.20 | 0.04 | 1.28 |
| **H06 (0-2 cm)** | 405 | -21,7398 | -40,0887 | 0 | 1.42 | 8.77 | 46.06 | 3.04 | 134.60 | 0.70 | 1.07 | 1.52 | 0.16 | 6.50 |
| **H07 (0-2 cm)** | 701 | -21,6873 | -40,0394 | 0 | 0.78 | 12.03 | 40.18 | 3.24 | 84.42 | 1.49 | 1.84 | 3.34 | 1.65 | 11.93 |
| **H08 (0-2 cm)** | 1006 | -21,6719 | -39,9688 | 22 | 1.07 | 13.27 | 37.70 | 2.71 | 63.60 | 0.79 | 1.61 | 2.35 | 0.27 | 7.41 |
| **H09 (0-2 cm)** | 1302 | -21,6560 | -39,8997 | 0 | 0.64 | 11.67 | 37.39 | 1.41 | 31.20 | 1.02 | 1.95 | 4.67 | 0.21 | 10.39 |
| **H10 (0-2 cm)** | 1900 | -21,6212 | -39,5964 | 0 | 0.67 | 7.60 | 54.65 | 0.51 | 29.90 | 0.15 | 0.17 | 0.15 | 0.04 | 0.91 |
| **H11 (0-2 cm)** | 2434 | -21,6221 | -39,0511 | 0 | 0.23 | 6.63 | 75.02 | 0.16 | 16.60 | 0.18 | 0.15 | 0.20 | 0.04 | 1.15 |

# APPENDIX A

Continued

| Station | Depth (m) | Lat. (WGS84) | Long. (WGS84) | N (ind./50cm$^3$) | Biomass (mg C.g$^{-1}$) | TOC (mg.g$^{-1}$) | $CaCO_3$ (%) | Phytopigment (µg.g$^{-1}$) | Flux (mgC.m$^{-2}$.day$^{-1}$) | Terr (µg.g$^{-1}$) | Zoo (µg.g$^{-1}$) | PP (µg.g$^{-1}$) | Bacteria (µg.g$^{-1}$) | Lipids (µg.g$^{-1}$) |
|---|---|---|---|---|---|---|---|---|---|---|---|---|---|---|
| **H12 (0-2 cm)** | 2953 | -21,6105 | -38,5410 | 0 | 0.20 | 2.50 | 72.93 | 0.05 | 14.20 | 0.02 | 0.01 | 0.05 $10^{-1}$ | * | 0.28 |
| **I06 (0-2 cm)** | 417 | -21,2278 | -40,2499 | 0 | 2.27 | 18.43 | 43.12 | 8.52 | 105.86 | 1.67 | 2.81 | 6.40 | 0.51 | 18.63 |
| **I07 (0-2 cm)** | 682 | -21,1872 | -40,2148 | 0 | 0.77 | 15.33 | 39.84 | 3.82 | 51.74 | 1.65 | 2.53 | 4.60 | 0.30 | 13.74 |
| **I08 (0-2 cm)** | 993 | -21,1851 | -40,1534 | 15 | 0.53 | 13.93 | 36.15 | 2.22 | 34.37 | 1.18 | 1.69 | 3.13 | 0.22 | 9.90 |
| **I09 (0-2 cm)** | 1300 | -21,1858 | -40,0523 | 5 | 1.10 | 11.37 | 34.61 | 2.44 | 27.66 | 0.94 | 1.03 | 1.67 | 0.16 | 5.90 |
| **I10 (0-2 cm)** | 1879 | -21,1842 | -39,6625 | 0 | 0.64 | 7.37 | 53.06 | 0.49 | 16.31 | 0.48 | 0.29 | 0.29 | 0.05 | 1.73 |
| **I11 (0-2 cm)** | 2407 | -21,1881 | -39,0850 | 0 | 0.21 | 4.33 | 77.12 | 0.18 | 13.21 | 0.42 | 0.36 | 0.36 | 0.10 | 3.02 |
| **I12 (0-2 cm)** | 3110 | -21,1872 | -38,4494 | 0 | 0.16 | 2.47 | 83.23 | 0.08 | 11.17 | 0.13 | 0.17 | 0.29 | 0.03 | 1.35 |
| **I06 (0-1 cm)** | 417 | -21,2278 | -40,2499 | 0 | * | * | * | * | * | * | * | * | * | * |
| **I06 (1-2 cm)** | 417 | -21,2278 | -40,2499 | 0 | * | * | * | * | * | * | * | * | * | * |
| **I06 (2-3 cm)** | 417 | -21,2278 | -40,2499 | 0 | * | * | * | * | * | * | * | * | * | * |
| **I06 (3-4 cm)** | 417 | -21,2278 | -40,2499 | 0 | * | * | * | * | * | * | * | * | * | * |
| **I06 (4-5 cm)** | 417 | -21,2278 | -40,2499 | 0 | * | * | * | * | * | * | * | * | * | * |
| **I06 (5-6 cm)** | 417 | -21,2278 | -40,2499 | 0 | * | * | * | * | * | * | * | * | * | * |
| **I06 (6-7 cm)** | 417 | -21,2278 | -40,2499 | 0 | * | * | * | * | * | * | * | * | * | * |
| **I06 (7-8 cm)** | 417 | -21,2278 | -40,2499 | 0 | * | * | * | * | * | * | * | * | * | * |
| **I06 (8-9 cm)** | 417 | -21,2278 | -40,2499 | 0 | * | * | * | * | * | * | * | * | * | * |

## APPENDIX A

Continued

| Station | Depth (m) | Lat. (WGS84) | Long. (WGS84) | N (ind./50cm$^3$) | Biomass (mg C.g$^{-1}$) | TOC (mg.g$^{-1}$) | $CaCO_3$ (%) | Phytopigment (µg.g$^{-1}$) | Flux (mgC.m$^{-2}$.day$^{-1}$) | Terr (µg.g$^{-1}$) | Zoo (µg.g$^{-1}$) | PP (µg.g$^{-1}$) | Bacteria (µg.g$^{-1}$) | Lipids (µg.g$^{-1}$) |
|---|---|---|---|---|---|---|---|---|---|---|---|---|---|---|
| **I06 (9-10 cm)** | 417 | -21,2278 | -40,2499 | 0 | * | * | * | * | * | * | * | * | * | * |
| **I08 (0-1 cm)** | 417 | -21,2278 | -40,2499 | 16 | * | * | * | * | * | * | * | * | * | * |
| **I08 (1-2 cm)** | 993 | -21,1851 | -40,1534 | 12 | * | * | * | * | * | * | * | * | * | * |
| **I08 (2-3 cm)** | 993 | -21,1851 | -40,1534 | 12 | * | * | * | * | * | * | * | * | * | * |
| **I08 (3-4 cm)** | 993 | -21,1851 | -40,1534 | 0 | * | * | * | * | * | * | * | * | * | * |
| **I08 (4-5 cm)** | 993 | -21,1851 | -40,1534 | 6 | * | * | * | * | * | * | * | * | * | * |
| **I08 (5-6 cm)** | 993 | -21,1851 | -40,1534 | 0 | * | * | * | * | * | * | * | * | * | * |
| **I08 (6-7 cm)** | 993 | -21,1851 | -40,1534 | 0 | * | * | * | * | * | * | * | * | * | * |
| **I08 (7-8 cm)** | 993 | -21,1851 | -40,1534 | 0 | * | * | * | * | * | * | * | * | * | * |
| **I08 (8-9 cm)** | 993 | -21,1851 | -40,1534 | 0 | * | * | * | * | * | * | * | * | * | * |
| **I08 (9-10 cm)** | 993 | -21,1851 | -40,1534 | 0 | * | * | * | * | * | * | * | * | * | * |

## Appendix A. Supplementary data

Geographical coordinates of the analyzed stations (WGS84), water depth, and density of living specimens of *R. pyriformis* n. sp. (number of individuals/50cm$^3$) and biotic and abiotic variables. Where: Biomass= Bacterial biomass (mg C.g$^{-1}$); TOC= Total organic carbon (mg.g$^{-1}$); $CaCO_3$=Calcium carbonate content (%); Phytopigment (µg.g$^{-1}$), Flux=Vertical particulate organic flux (mgC.m$^{-2}$.day$^{-1}$); Terr= Alloctone and/or continental derivative biomarker (µg.g$^{-1}$); Zoo= Zooplankton and/or fauna biomarker (µg.g$^{-1}$); PP= Phytoplankton and/or primary producers biomarker (µg.g$^{-1}$); Bacteria biomarker (µg.g$^{-1}$); Lipids (total lipids; µg.g$^{-1}$)*= no data.

| Station | Depth (m) | Lat. (WGS84) | Long. (WGS84) | N (ind./50cm$^3$) | Biomass (mg C.g$^{-1}$) | TOC (mg.g$^{-1}$) | $CaCO_3$ (%) | Phytopigment (µg.g$^{-1}$) | Flux (mgC.m$^{-2}$.day$^{-1}$) | Terr (µg.g$^{-1}$) | Zoo (µg.g$^{-1}$) | PP (µg.g$^{-1}$) | Bacteria (µg.g$^{-1}$) | Lipids (µg.g$^{-1}$) |
|---|---|---|---|---|---|---|---|---|---|---|---|---|---|---|
| **A06 (0-2 cm)** | 391 | -23,6331 | -41,3288 | 0 | 1.70 | 7.37 | 35.30 | 4.03 | 38.64 | 0.66 | 1.56 | 2.54 | 0.29 | 8.11 |
| **A07 (0-2 cm)** | 699 | -23,6560 | -41,3083 | 0 | 1.51 | 15.63 | 31.00 | 3.83 | 38.47 | 1.41 | 3.08 | 4.01 | 0.42 | 14.71 |
| **A08 (0-2 cm)** | 1018 | -23,6860 | -41,2684 | 5 | 0.74 | 14.33 | 30.59 | 3.65 | 27.20 | 1.21 | 2.70 | 5.16 | 0.45 | 19.24 |
| **A09 (0-2 cm)** | 1319 | -23,7527 | -41,1981 | 2 | 1.06 | 10.43 | 33.44 | 2.69 | 20.27 | 0.42 | 0.94 | 1.63 | 0.26 | 10.40 |
| **A10 (0-2 cm)** | 1935 | -23,8664 | -41,0792 | 0 | 0.47 | 9.80 | 42.18 | 0.57 | 10.46 | 0.34 | 0.64 | 0.60 | 0.15 | 6.44 |
| **A11 (0-2 cm)** | 2493 | -24,0239 | -40,9043 | 0 | 0.23 | 6.67 | 56.81 | 0.17 | 9.33 | 0.44 | 0.49 | 0.55 | 0.20 | 3.65 |
| **A12 (0-2 cm)** | 3035 | -24,4900 | -40,3902 | 0 | 0.19 | 5.20 | 71.03 | 0.15 | 9.27 | 0.17 | 0.20 | 0.16 | 0.07 | 1.32 |
| **B06 (0-2 cm)** | 412 | -23,1734 | -40,9471 | 0 | 0.73 | 20.50 | 31.16 | 5.92 | 90.54 | 2.11 | 2.60 | 5.47 | 0.60 | 18.11 |
| **B07 (0-2 cm)** | 738 | -23,2176 | -40,9609 | 0 | 1.60 | 20.73 | 31.15 | 10.19 | 43.79 | 1.98 | 6.54 | 9.33 | 0.63 | 27.17 |
| **B08 (0-2 cm)** | 1001 | -23,2307 | -40,9320 | 11 | 0.94 | 16.30 | 31.29 | 4.78 | 26.71 | 1.81 | 3.31 | 7.53 | 0.43 | 19.63 |
| **B09 (0-2 cm)** | 1228 | -23,2537 | -40,8986 | 4 | 1.03 | 15.80 | 29.13 | 3.16 | 22.05 | 1.74 | 1.72 | 3.29 | 0.47 | 14.01 |

## APPENDIX A

Continued

| Station | Depth (m) | Lat. (WGS84) | Long. (WGS84) | N (ind./50cm$^3$) | Biomass (mg C.g$^{-1}$) | TOC (mg.g$^{-1}$) | $CaCO_3$ (%) | Phytopigment (µg.g$^{-1}$) | Flux (mgC.m$^{-2}$.day$^{-1}$) | Terr (µg.g$^{-1}$) | Zoo (µg.g$^{-1}$) | PP (µg.g$^{-1}$) | Bacteria (µg.g$^{-1}$) | Lipids (µg.g$^{-1}$) |
|---|---|---|---|---|---|---|---|---|---|---|---|---|---|---|
| **B10 (0-2 cm)** | 1900 | -23,3104 | -40,7914 | 0 | 0.62 | 11.80 | 39.07 | 1.02 | 15.33 | 0.81 | 0.92 | 1.41 | 0.11 | 4.91 |
| **B11 (0-2 cm)** | 2492 | -23,4226 | -40,5998 | 0 | 0.26 | 7.53 | 58.93 | 0.26 | 11.26 | 0.49 | 0.54 | 0.69 | 0.20 | 3.96 |
| **B12 (0-2 cm)** | 2424 | -23,7556 | -39,9999 | 0 | 0.28 | 4.40 | 58.48 | 0.10 | 9.39 | 0.14 | 0.09 | 0.12 | 0.01 | 0.70 |
| **D06 (0-2 cm)** | 401 | -22,5603 | -40,4442 | 0 | 0.76 | 10.63 | 36.77 | 2.94 | 44.90 | 1.17 | 1.61 | 2.47 | 0.29 | 9.14 |
| **D07 (0-2 cm)** | 696 | -22,6075 | -40,3756 | 0 | 0.66 | 11.70 | 34.38 | 2.34 | 32.10 | 0.84 | 1.08 | 1.76 | 0.21 | 6.22 |
| **D08 (0-2 cm)** | 1010 | -22,6829 | -40,2942 | 7 | 0.44 | 13.20 | 36.03 | 1.53 | 28.80 | 1.43 | 2.50 | 3.99 | 0.32 | 12.68 |
| **D10 (0-2 cm)** | 1921 | -22,8231 | -40,1386 | 0 | 0.20 | 7.23 | 48.41 | 0.72 | 13.00 | 0.23 | 0.52 | 0.68 | 0.10 | 2.43 |
| **D11 (0-2 cm)** | 2492 | -22,8713 | -40,0865 | 0 | 0.22 | 3.25 | 65.48 | 0.64 | 11.90 | 0.31 | 0.26 | 0.41 | 0.04 | 1.47 |
| **D12 (0-2 cm)** | 3016 | -23,3101 | -39,5992 | 0 | 0.19 | 4.17 | 73.86 | 0.14 | 9.10 | 0.28 | 0.20 | 0.20 | 0.04 | 1.28 |
| **H06 (0-2 cm)** | 405 | -21,7398 | -40,0887 | 0 | 1.42 | 8.77 | 46.06 | 3.04 | 134.60 | 0.70 | 1.07 | 1.52 | 0.16 | 6.50 |
| **H07 (0-2 cm)** | 701 | -21,6873 | -40,0394 | 0 | 0.78 | 12.03 | 40.18 | 3.24 | 84.42 | 1.49 | 1.84 | 3.34 | 1.65 | 11.93 |
| **H08 (0-2 cm)** | 1006 | -21,6719 | -39,9688 | 22 | 1.07 | 13.27 | 37.70 | 2.71 | 63.60 | 0.79 | 1.61 | 2.35 | 0.27 | 7.41 |
| **H09 (0-2 cm)** | 1302 | -21,6560 | -39,8997 | 0 | 0.64 | 11.67 | 37.39 | 1.41 | 31.20 | 1.02 | 1.95 | 4.67 | 0.21 | 10.39 |
| **H10 (0-2 cm)** | 1900 | -21,6212 | -39,5964 | 0 | 0.67 | 7.60 | 54.65 | 0.51 | 29.90 | 0.15 | 0.17 | 0.15 | 0.04 | 0.91 |
| **H11 (0-2 cm)** | 2434 | -21,6221 | -39,0511 | 0 | 0.23 | 6.63 | 75.02 | 0.16 | 16.60 | 0.18 | 0.15 | 0.20 | 0.04 | 1.15 |

## APPENDIX A

Continued

| Station | Depth | Lat. | Long. | N | Biomass | TOC | $CaCO_3$ | Phytopigment | Flux | Terr | Zoo | PP | Bacteria | Lipids |
|---|---|---|---|---|---|---|---|---|---|---|---|---|---|---|
| | (m) | (WGS84) | (WGS84) | (ind./50cm$^3$) | (mg C.g$^{-1}$) | (mg.g$^{-1}$) | (%) | (µg.g$^{-1}$) | (mgC.m$^{-2}$.day$^{-1}$) | (µg.g$^{-1}$) | (µg.g$^{-1}$) | (µg.g$^{-1}$) | (µg.g$^{-1}$) | (µg.g$^{-1}$) |
| **H12 (0-2 cm)** | 2953 | -21,6105 | -38,5410 | 0 | 0.20 | 2.50 | 72.93 | 0.05 | 14.20 | 0.02 | 0.01 | 0.05 $10^{-1}$ | * | 0.28 |
| **I06 (0-2 cm)** | 417 | -21,2278 | -40,2499 | 0 | 2.27 | 18.43 | 43.12 | 8.52 | 105.86 | 1.67 | 2.81 | 6.40 | 0.51 | 18.63 |
| **I07 (0-2 cm)** | 682 | -21,1872 | -40,2148 | 0 | 0.77 | 15.33 | 39.84 | 3.82 | 51.74 | 1.65 | 2.53 | 4.60 | 0.30 | 13.74 |
| **I08 (0-2 cm)** | 993 | -21,1851 | -40,1534 | 15 | 0.53 | 13.93 | 36.15 | 2.22 | 34.37 | 1.18 | 1.69 | 3.13 | 0.22 | 9.90 |
| **I09 (0-2 cm)** | 1300 | -21,1858 | -40,0523 | 5 | 1.10 | 11.37 | 34.61 | 2.44 | 27.66 | 0.94 | 1.03 | 1.67 | 0.16 | 5.90 |
| **I10 (0-2 cm)** | 1879 | -21,1842 | -39,6625 | 0 | 0.64 | 7.37 | 53.06 | 0.49 | 16.31 | 0.48 | 0.29 | 0.29 | 0.05 | 1.73 |
| **I11 (0-2 cm)** | 2407 | -21,1881 | -39,0850 | 0 | 0.21 | 4.33 | 77.12 | 0.18 | 13.21 | 0.42 | 0.36 | 0.36 | 0.10 | 3.02 |
| **I12 (0-2 cm)** | 3110 | -21,1872 | -38,4494 | 0 | 0.16 | 2.47 | 83.23 | 0.08 | 11.17 | 0.13 | 0.17 | 0.29 | 0.03 | 1.35 |
| **I06 (0-1 cm)** | 417 | -21,2278 | -40,2499 | 0 | * | * | * | * | * | * | * | * | * | * |
| **I06 (1-2 cm)** | 417 | -21,2278 | -40,2499 | 0 | * | * | * | * | * | * | * | * | * | * |
| **I06 (2-3 cm)** | 417 | -21,2278 | -40,2499 | 0 | * | * | * | * | * | * | * | * | * | * |
| **I06 (3-4 cm)** | 417 | -21,2278 | -40,2499 | 0 | * | * | * | * | * | * | * | * | * | * |
| **I06 (4-5 cm)** | 417 | -21,2278 | -40,2499 | 0 | * | * | * | * | * | * | * | * | * | * |
| **I06 (5-6 cm)** | 417 | -21,2278 | -40,2499 | 0 | * | * | * | * | * | * | * | * | * | * |
| **I06 (6-7 cm)** | 417 | -21,2278 | -40,2499 | 0 | * | * | * | * | * | * | * | * | * | * |
| **I06 (7-8 cm)** | 417 | -21,2278 | -40,2499 | 0 | * | * | * | * | * | * | * | * | * | * |
| **I06 (8-9 cm)** | 417 | -21,2278 | -40,2499 | 0 | * | * | * | * | * | * | * | * | * | * |

# APPENDIX A

Continued

| Station | Depth (m) | Lat. (WGS84) | Long. (WGS84) | N (ind./50cm$^3$) | Biomass (mg C.g$^{-1}$) | TOC (mg.g$^{-1}$) | $CaCO_3$ (%) | Phytopigment (µg.g$^{-1}$) | Flux (mgC.m$^{-2}$.day$^{-1}$) | Terr (µg.g$^{-1}$) | Zoo (µg.g$^{-1}$) | PP (µg.g$^{-1}$) | Bacteria (µg.g$^{-1}$) | Lipids (µg.g$^{-1}$) |
|---|---|---|---|---|---|---|---|---|---|---|---|---|---|---|
| **I06 (9-10 cm)** | 417 | -21,2278 | -40,2499 | 0 | * | * | * | * | * | * | * | * | * | * |
| **I08 (0-1 cm)** | 417 | -21,2278 | -40,2499 | 16 | * | * | * | * | * | * | * | * | * | * |
| **I08 (1-2 cm)** | 993 | -21,1851 | -40,1534 | 12 | * | * | * | * | * | * | * | * | * | * |
| **I08 (2-3 cm)** | 993 | -21,1851 | -40,1534 | 12 | * | * | * | * | * | * | * | * | * | * |
| **I08 (3-4 cm)** | 993 | -21,1851 | -40,1534 | 0 | * | * | * | * | * | * | * | * | * | * |
| **I08 (4-5 cm)** | 993 | -21,1851 | -40,1534 | 6 | * | * | * | * | * | * | * | * | * | * |
| **I08 (5-6 cm)** | 993 | -21,1851 | -40,1534 | 0 | * | * | * | * | * | * | * | * | * | * |
| **I08 (6-7 cm)** | 993 | -21,1851 | -40,1534 | 0 | * | * | * | * | * | * | * | * | * | * |
| **I08 (7-8 cm)** | 993 | -21,1851 | -40,1534 | 0 | * | * | * | * | * | * | * | * | * | * |
| **I08 (8-9 cm)** | 993 | -21,1851 | -40,1534 | 0 | * | * | * | * | * | * | * | * | * | * |
| **I08 (9-10 cm)** | 993 | -21,1851 | -40,1534 | 0 | * | * | * | * | * | * | * | * | * | * |